\documentclass[aps,prl,showpacs,amsmath,twocolumn,amssymb,superscriptaddress,letterpaper]{revtex4}
\usepackage{times}
\usepackage{amsfonts}
\usepackage{mathrsfs}
\usepackage{graphicx}
\usepackage{dcolumn}
\usepackage{bm}
\usepackage{color}

\usepackage[colorlinks,bookmarks=false,citecolor=blue,linkcolor=red,urlcolor=blue]{hyperref}
\usepackage{bibunits}

\def\Re {\mbox{Re}}
\def\Im {\mbox{Im}}
\def\be{\begin{equation}}       \def\ee{\end{equation}}
\def\bea{\begin{eqnarray}}      \def\eea{\end{eqnarray}}

\begin{document}

\begin{bibunit}

\title{Non-Hermitian Hopf-Link Exceptional Line Semimetals}
\author{Zhesen Yang}
\affiliation{Beijing National Laboratory for Condensed Matter Physics,
and Institute of Physics, Chinese Academy of Sciences, Beijing 100190, China}
\affiliation{University of Chinese Academy of Sciences, Beijing 100049, China}

\author{Jiangping Hu}\email{jphu@iphy.ac.cn}
\affiliation{Beijing National Laboratory for Condensed Matter Physics,
and Institute of Physics, Chinese Academy of Sciences, Beijing 100190, China}
\affiliation{CAS Center of Excellence in Topological Quantum Computation and Kavli Institute of Theoretical Sciences, University of Chinese Academy of Sciences,
Beijing, 100190, China}
\affiliation{Collaborative Innovation Center of Quantum Matter, Beijing, 100871, China}

\date{\today}

\begin{abstract}
We study a new class of non-Hermitian topological phases in three dimension in the absence of any symmetry, where the topological robust band degeneracies are Hopf-link exceptional lines. As a concrete example, we investigate the non-Hermitian band structures of nodal line semimetals under non-Hermitian perturbations, where the Fermi surfaces can transit from $1d$ nodal lines to $2d$ twisting surfaces with Hopf-link boundaries when the winding number defined along the nodal line is $\pm1$. The linking numbers of these linked exceptional line phases are also proposed, based on the integral of Chern-Simons form over the Brillouin zone.  
\end{abstract}


\maketitle

{\em Introduction}---Non-Hermitian Hamiltonians have been widely utilized in physical systems\cite{Bender1998, Bender2007,book}, such as open quantum systems\cite{Rotter2009}, wave systems with gain and loss\cite{Regensburger2012,Ruter2010,Lin2011,Feng2013,Guo2009,Peng2014,Hodaei2017,Hodaei2014,Feng2014,Gao2015,Xu2016,Ashida2017,Chen2017,Ding2016,Ganainy2018,Longhi2018}
         and interacting electron systems\cite{Fu2017,Fu2018a,Fu2018b}. A Hamiltonian describing these systems  generally includes three parts\cite{Rotter2009}: the intrinsic part $H_S$, the external environment $H_E$, and the coupling between them $H_{SE}$.  By tracing out the environment degree, an effective Hamiltonian, $H_{S}^{eff}$,  can be obtained\cite{Rotter2009}. The non-Hermitian terms in $H_{S}^{eff}$ can be viewed as a dynamical instability that is imposed to $H_S$  through complex  eigenenergies whose imaginary parts specify lifetimes. Many intriguing phenomena have been proposed for non-Hermitian systems, such as unidirectional invisibility\cite{Lin2011}, single-mode lasers\cite{Feng2014, Hodaei2014}, enhanced sensitivity in optics\cite{Chen2017, Hodaei2017},  bulk Fermi-arc\cite{Fu2017,Fu2018a} and non-Hermitian skin effect\cite{Wang2018a,Wang2018b}. 

The concept of topological invariants has been proposed to classify both gapped and gapless phases in Hermitian systems\cite{Kane2010, Zhang2011,Fang2016,Ryu2016,Weyl}, and also has been applied to  non-Hermitian systems recently\cite{Shen2018,Esaki2011,Lee,Nori,Lieu,menke,Runder,Song,Xiong,Torrest,Gong, Wang2018a, Wang2018b,Huang,Hu,Gong2,Runder2,Zhu,Gong3,Wang2,Ueda,Ni,Zyuzin,Fan,Klett,Zhou2,Yuce,Hu2,Duan, Ke,Zeuner, Xiao2,Wm,Poli, Parto, Zhao3, Zhan3, Avila, Wei Yi, Hui Zhai, Ueda new,Timothy, Alvarez,Kunst,González2018,González2017}. One crucial observation is that topologically stable band degeneracies in the non-Hermitian case only require tuning two independent parameters instead of three ones required in the Hermitian case\cite{Weyl, Shen2018}.   Thus in two dimensional ($2d$) non-Hermitian models, the exceptional points (EPs), which can not be eliminated by small perturbations in the absence of any symmetry,  act in a same way as Weyl points in three dimensional ($3d$) Hermitian systems\cite{Shen2018}.   In $3d$, the stable degeneracies in non-Hermitian systems must be one dimensional exceptional lines (ELs), which can be linked or knotted in a nontrivial way. This is similar to five dimensional ($5d$) Weyl semimetals\cite{Lian1}, where the Weyl points are generalized to the Weyl surfaces and the topological invariants are related to the linking number. Based on this observation, new topological invariants must be required to describe the linking structure of the ELs. 

In the Hermitian case, the  linking degeneracies have already been studied theoretically, such as nodal link and nodal knot semimetals\cite{XQSun,Lian1,Chen4,Wang link,Yee, Ezawa,Wang knot, Hasan, Wan4,Gong5,Zhang5}. However, the proposed models have not been realized experimentally due to the requirement of having significant and suitable hoppings beyond the nearest neighbor, which is not immediately available in materials.  Although it has been proposed to realize nodal link or knot semimetals from a nodal chain semimetal by adding some particular perturbations,  the transition requires  the touching point of nodal lines to satisfy strict quadratic dispersion along all directions\cite{Yangzs} which is forbidden by the constrain of mirror or glade plane symmetries\cite{Fang2016}.  Therefore, it is natural to generalize and investigate linking degeneracies in non-Hermitian systems.

\begin{figure}[t]
\centerline{\includegraphics[height=3.6cm]{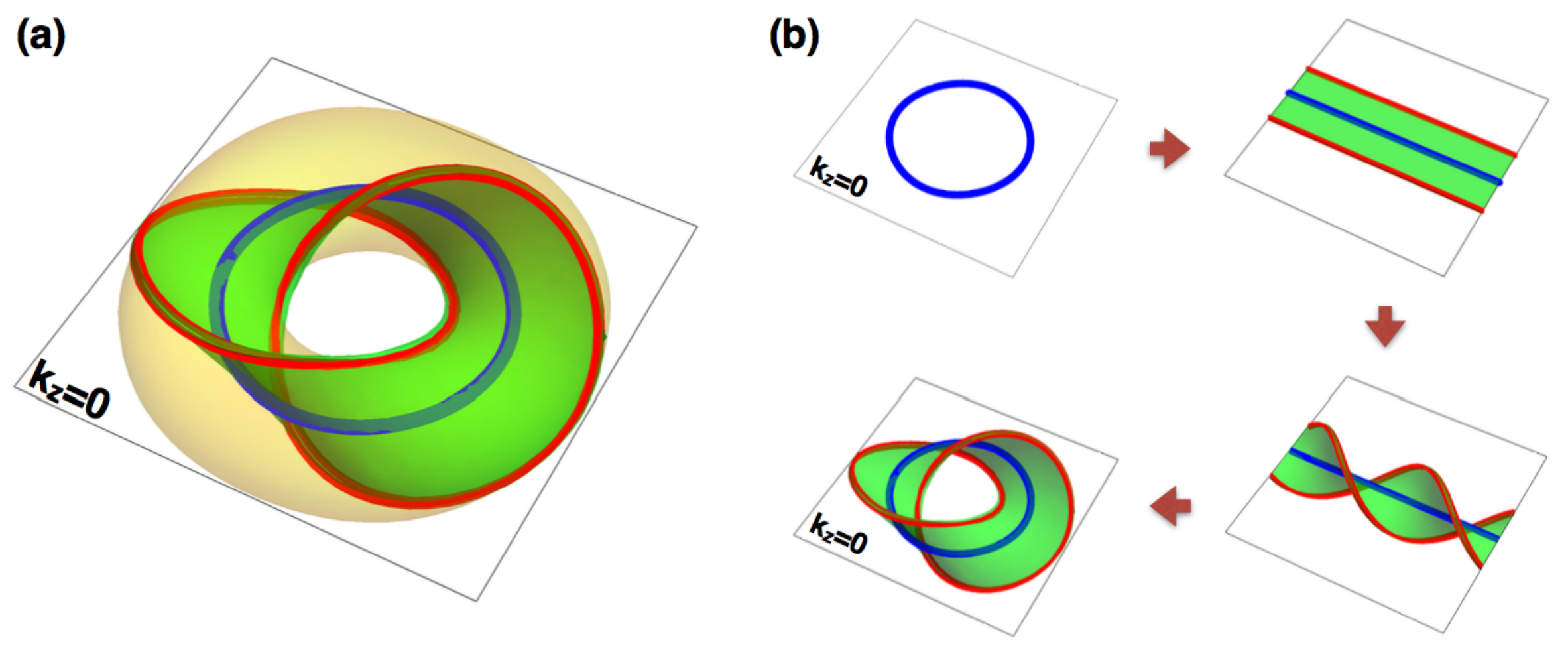}}
\caption{A illustration of forming Hopf-link exceptional line (red line) semimetal. Starting from a nodal line (blue line) semimetal, the Fermi surface (FS) will be regulated by Eq. (\ref{E4}). (a). The yellow surface represents the condition $|{\bf h}|^2-\lambda^2|{\bf g}|^2\leq 0$ and the green one represents the regulated FS (with a finite life time), whose boundaries (red curves) are Hopf-link exceptional lines  while the original nodal line (blue) is on the green FS. (b) shows the twisting and linking structure. 
\label{F1}}
\end{figure}

In this paper, we propose a simple method to realize the Hopf-link EL semimetals by non-Hermitian perturbations in a nodal line semimetal. It is found that the Fermi surfaces (FSs)  transit from $1d$ nodal lines to $2d$ twisting FSs with linked $1d$ boundaries, as shown in FIG. \ref{F1} (a) with blue, green and red ones respectively. Due to the vanishing of the imaginary part of the eigenvalues of the Hamiltonian, only these boundaries, known as ELs,  are dynamically stable. In order to describe these nontrivial linking ELs,  we construct a new topological invariant (linking number), based on the integral of the Chern-Simons form over the Brillouin zone (BZ) in a duality Hermitian Hamiltonian. For  the nodal line semimetals, the linking structures of the ELs can be captured by a winding number along the nodal line defined by the non-Hermitian perturbations.         

{\em Non-Hermitian semimetals---}In the Hermitian case, the robust band degeneracies require tuning three parameters in the absence of any symmetry\cite{Weyl}. As a result, the Weyl points in three dimension are stable and cannot be eliminated by small perturbations, whereas the Dirac points in two dimension and nodal lines in three dimension are unstable in the absence of symmetry. However, in the non-Hermitian case, only two parameters are required to obtain the stable band degeneracies\cite{Shen2018}. This can be understood by the following two band non-Hermitian model, whose Bloch Hamiltonian is,      
\begin{equation}
\begin{aligned}
&\mathcal{H}({\bm k})=[\bm{h}({\bm k})+i\lambda\bm{g}({\bm k})]\cdot\bm{\sigma},
\label{E2}\end{aligned}
\end{equation}
         where $\bm{f}=(f_x,f_y,f_z)$ with $\bm{f}=\bm{h},\bm{g},\bm{\sigma},\bm{k}$, and $\lambda$ is the energy scale of non-Hermitian terms. The eigenvalues of the above Hamiltonian are, 
\begin{equation}
\epsilon_\pm ({\bm k})=\pm \sqrt{(|{\bm h}|^2-\lambda^2|{\bm g}|^2)+2i\lambda {\bm h}\cdot {\bm g} }. 
\label{n1}
\end{equation}
         The condition for the gap closing is $\epsilon_+ ({\bm k})=\epsilon_- ({\bm k})=0$, which implies 
\begin{equation}
|{\bm h}({\bm k})|^2-\lambda^2|{\bm g}({\bm k})|^2=0, \quad 2\lambda {\bm h}({\bm k})\cdot {\bm g}({\bm k})=0.
\label{n2}
\end{equation}
         In contrast to the Hermitian system, there are only two equations for the emergence of robust band degeneracies. As a consequence, in three dimension, the solution of the above equations is in general a $1d$ manifold (ELs), which are robust and cannot be eliminated by weak perturbations in the absence of any symmetry.  

As interpreted by Ref. \cite{book}, the real and imaginary parts of the eigenvalues $\epsilon_\pm ({\bm k})$, are related to the energies and inverse of lifetime of the system, $E_{\pm}({\bm k})=\Re[\epsilon_\pm ({\bm k})]$ and $\quad \frac{1}{2}\Gamma_{\pm}({\bm k})=\Im[\epsilon_\pm ({\bm k})]$. One should notice that the $\pm$ sign of $E_{\pm}({\bm k})$ are not necessary corresponding to the occupied or unoccupied states. However, for convenience, we can still define the FSs  as the vanishing of the real part of the eigenvalues $\epsilon_{\pm}({\bm k})$, which are determined by the following two equations, 
\begin{equation}
|{\bm h}({\bm k})|^2-\lambda^2|{\bm g}({\bm k})|^2\leq0, \quad 2\lambda {\bm h}({\bm k})\cdot {\bm g}({\bm k})=0.
\label{E4}
\end{equation}
The first equation $|{\bm h}({\bm k})|^2-\lambda^2|{\bm g}({\bm k})|^2\leq0$ with "$=$" determines a $2d$ closed surface, and "$<$" extends its region  in the BZ, as shown in FIG. \ref{F1}(a) with the yellow one. Thus on the one hand, the set of "${\bm k}$" points satisfying Eq. (\ref{E4}) with "$=$" is equivalent to the points satisfying Eq. (\ref{n2}). These are the  
         $1d$ ELs, which are dynamically stable due to the vanishing of the imaginary part $\Gamma_{\pm}({\bm k})$. On the other hand, the Eq. (\ref{E4}) with "$\leq$"  determines a $2d$ FSs, whose boundaries are the ELs, as shown in FIG. \ref{F1}(a) with green and red ones. Similar to the nodal link and nodal knot semimetals\cite{Chen4,Wang link,Yee, Ezawa,Wang knot, Hasan, Wan4,Gong5,Zhang5,Yangzs}, the ELs can also be linked together, or form some nontrivial knots in the absence of any symmetry. Next we will define the linking number for these novel EL semimetals. 

\begin{figure}[b]
\centerline{\includegraphics[height=3cm]{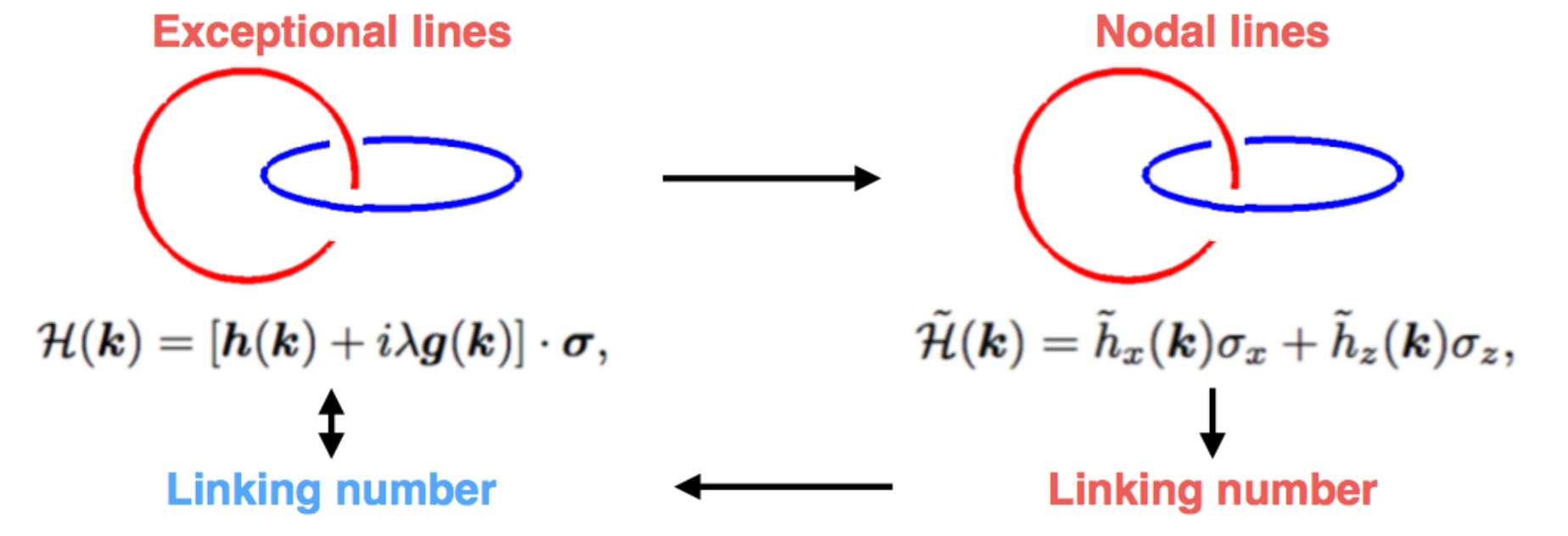}}
\caption{The strategy for the definition of linking number in  exceptional line (EL) semimetals, e.g. Hopf-link EL semimetal. We first map the non-Hermitian model Eq. (\ref{E2}) to a Hermitian nodal line model Eq. (\ref{n3}), whose band degeneracies have the same form. As a result, the ELs and nodal lines share the same linking number, which can be calculated from the Hermitian nodal line model.
\label{FF2}}
\end{figure}

{\em{Topological invariants}---}Several topological invariants have been proposed for non-Hermitian systems, such as winding number\cite{Lee}, Chern number\cite{Shen2018, Duan} and vorticity\cite{Shen2018}. However they can not capture the linking or knotting structures of the ELs. Unlike the linking degeneracies in the Hermitian case which are protected by $PT$ symmetry (or sublattice symmetry), the ELs in the non-Hermitian systems are not protected by any symmetry and can not be gapped by adding small perturbations. This is similar to the $5d$ generalization of Weyl semimetals, where the basic topological invariant is the linking number of the Weyl surfaces\cite{Lian1}. This feature causes difficulty to define the linking number of the ELs as the Berry curvature can not be defined along these ELs.  Thus, a simple generalization of topological invariants from the Hermitian case\cite{Lian2} to the non-Hermitian case fails.

Our strategy to define the topological invariants of the EL semimetals with linking structures is to map the non-Hermitian Hamiltonian Eq. (\ref{E2}) to a Hermitian nodal line Hamiltonian, in which the parameter equations of nodal lines have the same form of ELs in the non-Hermitian system, as shown in Fig. \ref{FF2}. Observing that the ELs are determined by Eq. (\ref{n2}), the Hermitian nodal line Hamiltonian can be defined as 
\begin{equation}\begin{aligned}
\tilde{\mathcal{H}}({\bm k})=\tilde{h}_x({\bm k})&\sigma_x+\tilde{h}_z({\bm k})\sigma_z,\\
\tilde{h}_x({\bm k})=|{\bm h}({\bm k})|^2-\lambda^2|{\bm g}({\bm k})|^2&,\quad \tilde{h}_z({\bm k})=2\lambda {\bm h}({\bm k})\cdot {\bm g}({\bm k}).
\label{n3}\\
\end{aligned}\end{equation}
         One can check that the nodal lines determined by the above Hamiltonian $\tilde{\mathcal{H}}({\bm k})$ are $\tilde{h}_x({\bm k})=0$ and $\tilde{h}_z({\bm k})=0$, which are the same as Eq. (\ref{n2}). As a result, the nodal lines determined by Eq. (\ref{n3}) have the same form of ELs determined by Eq. (\ref{n2}), which means they have the same linking number. 

In contrast to the ELs, these nodal lines determined by Eq. (\ref{n3}) are protected by $PT$ symmetry (or sublattice symmetry) and can be gapped by adding a small $PT$ symmetric breaking (or sublattice symmetric breaking) term $m\sigma_y$, that is, 
\begin{equation}
\tilde{\mathcal{H}}({\bm k})=\tilde{h}_x({\bm k})\sigma_x+\tilde{h}_z({\bm k})\sigma_z+m\sigma_y. 
\label{n4}
\end{equation}
         Based on this term, we can define the Berry curvature $d{\bm a}$ in the whole BZ without singularity. Thus the linking number is captured by the integral of the Chern-Simons form over the BZ\cite{Lian2}, 
\begin{equation}
\frac{1}{4\pi}\int_{BZ}{\bm a}\wedge d{\bm a}=\pi\sum_{\alpha,\beta}\nu_\alpha\nu_\beta N(\mathcal{L}_\alpha,\mathcal{L}_\beta).
\label{n5}
\end{equation}
         Here $\alpha,\beta$ label different nodal lines $\mathcal{L}_\alpha$ and $\mathcal{L}_\beta$ determined by Eq. (\ref{n3}) in the BZ, e.g. the red and blue loop in Fig. \ref{FF2}, $\nu_\alpha$ and $\nu_\beta$ are the first $Z_2$ charge\cite{Fang2016} of the two nodal lines $\mathcal{L}_\alpha$ and $\mathcal{L}_\beta$, that is $\nu_{\alpha/\beta}=1/(2\pi)\oint_{\mathcal{C}_{\alpha/\beta}}{\bm a}({\bm k})\cdot d{\bm k}$, where $\mathcal{C}_{\alpha/\beta}$ are the loops enclosing the nodal lines $\mathcal{L}_\alpha$ and $\mathcal{L}_\beta$, ${\bm a}$ and $d{\bm a}$ are the Berry connection and Berry curvature of the Hamiltonian Eq. (\ref{n4}),  $N(\mathcal{L}_\alpha,\mathcal{L}_\beta)$ is the linking number of the nodal lines $\mathcal{L}_\alpha$ and $\mathcal{L}_\beta$, which have the same value of the ELs determined by Eq. (\ref{E2}). Like the $Z_2$ charge, the integral of the Chern-Simons form $1/(4\pi)\int_{BZ}{\bm a}\wedge d{\bm a}$ is only defined mod $2\pi$ due to gauge invariance. For a simple EL semimetal, e.g. the model proposed in Ref. \cite{Duan}, the integral of the Chern-Simons form $1/(4\pi)\int_{BZ}{\bm a}\wedge d{\bm a}$ is zero (where ${\bm a}$ and $d{\bm a}$ are defined in the duality Hermitian Hamiltonian Eq. (\ref{n3}) and Eq. (\ref{n4})). A simple non-trivial example is the Hopf-link EL semimetal, which will be discussed in the next section. 

{\em Hopf-link EL semimetals---} We show below the Hopf-link EL semimetal can be realized by adding non-Hermitian perturbations from a Hermitian nodal line semimetal. The Bloch Hamiltonian of the nodal line semimetal is given by
\begin{equation}\begin{aligned}
\mathcal{H}_{0}({\bm{k}})=h_x({\bm{k}})\sigma_x&+h_z({\bm{k}})\sigma_z,\\
h_x({\bm{k}})=\sin k_z,\quad h_z({\bm{k}})=m+&\cos k_x+\cos k_y+\cos k_z.
\label{E1}\\
\end{aligned}\end{equation}
         where $m$ is external parameter and $\sigma_x, \sigma_z$ are the Pauli matrices. Without losing generality, we have neglected the $\varepsilon_0({\bm{k}})\sigma_0$ term and have assumed $m=-21/8$ for the simplification of our discussion. In this case, the energy spectra  become $E_{0,\pm}({\bm k})=\pm\sqrt{h_x({\bm{k}})^2+h_z({\bm{k}})^2}$. The nodal line is determined by the following two equations $\sin k_z=0$ and $\cos k_x+\cos k_y=13/8$. In the continuum limit, the parameter equation of the nodal line can be approximated by $(k_x,k_y,k_z)=(\sqrt{3/4}\cos\theta,\sqrt{3/4}\sin\theta,0)$. An specific example of the nodal line is shown in Fig. \ref{F1} with the blue line. The topological invariant of this model is the first $Z_2$ charge\cite{Fang2016,Y.Chan}, which is defined by the integral of the Berry connection along a loop enclosing the nodal line. Because of the gauge invariance, the $\pm\pi$  represent one  state. The above nodal line semimetal also has a overlooked topological invariant---the linking number\cite{Lian2,XQSun}, even though its value is zero in this simple model. 

\begin{figure}[t]
\centerline{\includegraphics[height=2.7cm]{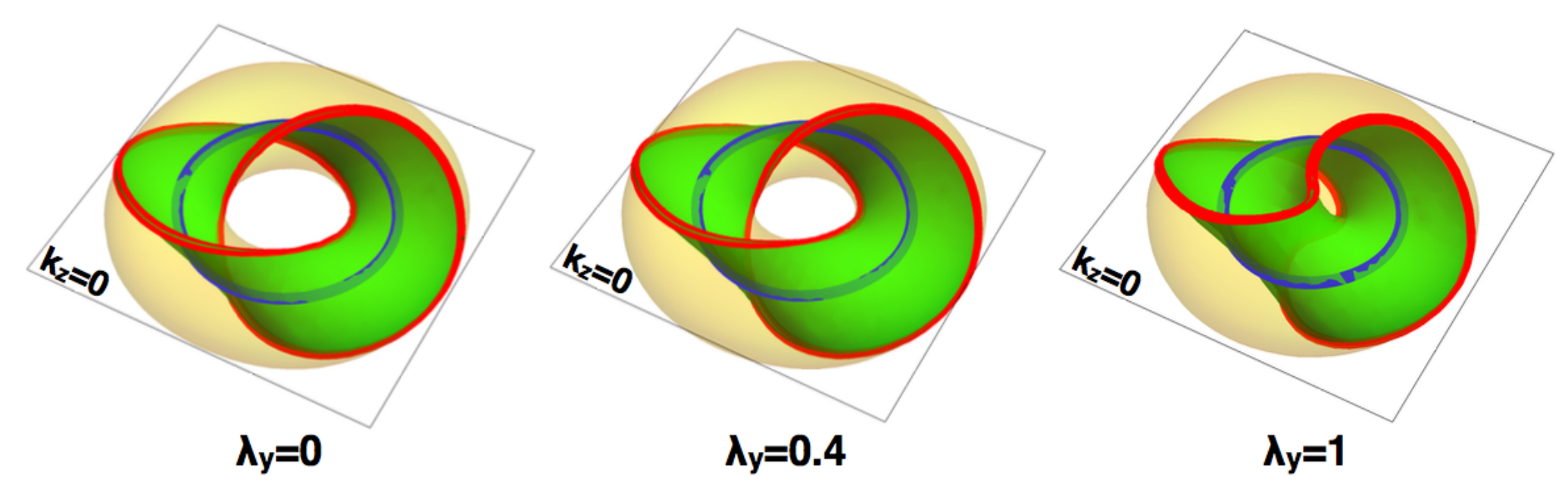}}
\caption{Exceptional lines (ELs) with different parameters of Eq. (\ref{E2}), (\ref{E1}) and (\ref{E5}). The red, blue, green and yellow color represent the ELs, nodal line, Fermi surface and $|\bm{h}|^2-\lambda^2|\bm{g}|^2\leq0$. One can notice the Hopf-link ELs are robust against weak sublattice symmetry breaking terms $\lambda_y$. The parameters in the calculation are $m=-21/8$ and $\lambda=1/2$.
\label{F5}}
\end{figure}

The non-Hermitian terms is chosen to be, 
\begin{equation}
g_{x}({\bm k})=\sin k_x,\quad g_y({\bm k})=\lambda_y,\quad g_z({\bm k})=\sin k_y,
\label{E5}
\end{equation}
         where $\lambda_y$ is a parameter that breaks the sublattice symmetry in the Hamiltonian (\ref{E2}) with Eq. (\ref{E1}) and (\ref{E5}). Fig. \ref{F5} shows the corresponding ELs with $\lambda=1/2$ and different values of $\lambda_y$. In order to simplify the discussion, we first assume $\lambda_y=0$. In the continuum limit around the $\Gamma$ point, the equation that determines the ELs is $(k_z+i\lambda k_x)^2+(-(k^2/2-3/8)+i\lambda k_y)^2=0$. This equation has two possible solutions $\pm i(k_z+i\lambda k_x)=-(k^2/2-3/8)+i\lambda k_y$, which correspond to the two ELs with the following parameter equations $(k_{x,\pm},k_{y,\pm},k_{z,\pm})=(\pm \lambda+\sqrt{(4\lambda^2+3)/4}\cos t$,$\sqrt{(4\lambda^2+3)/4(\lambda^2+1)}\sin t$, $\pm \lambda\sqrt{(4\lambda^2+3)/4(\lambda^2+1)}\sin t)$, where $t\in[-\pi,\pi]$. We can easily check these two loops  linked together when $\lambda\neq0$, because  the separation length between the two centers of the  loops, $2|\lambda|$, is lesser than the summation of the radius along the $x-$direction, $2\sqrt{(4\lambda^2+3)/4}$. We can also understand this linking structure based on the geometry and perturbation analysis. Notice the ELs are determined by Eq. (\ref{n2}), where the first equation $|{\bm h}({\bm k})|^2-\lambda^2|{\bm g}({\bm k})|^2\leq0$ acts as the $\bm{k}$ dependent doping level of the nodal line semimetals. When $0<\lambda\ll 1$, the geometry object determined by the above inequation is a solid torus  centered  at the nodal line when $|{\bm g}({\bm k}_{nl})|\neq0$. We can cut this nodal line and then straighten it with a periodic boundary condition as shown in Fig. \ref{F1}(b). Combining the second equation $\quad 2\lambda {\bm h}({\bm k})\cdot {\bm g}({\bm k})=0$, we can see that the intersection surface can be viewed as a strip with twisting boundary conditions. Fig. \ref{F1}(b) shows this procedure in which  the red boundary lines are the ELs linked together. If the twisting angle of the FS is $2\pi$, the boundary red lines just form a Hopf-link.
         When the sublattice symmetric breaking term is weak, that is $\lambda_y\ll1$, we can also argue Hopf-link structure is robust. Because $\lambda_y$ only modifies the first equation of Eq. (\ref{n2}), which render the a lagger torus (yellow one in Fig. \ref{F5}) around the original nodal line. While the twisting structure, which is determined by the second equation of Eq. (\ref{n2}), is not changed. As a result, the Hopf-link is quite robust to weak sublattice symmetric breaking terms. It should be emphasized that
         the geometry argument can only be applied for the nodal line semimetals with a small non-Hermitian perturbation because the solid torus transits to a solid sphere for a large $\lambda_y$. 
         
\begin{figure}[t]
\centerline{\includegraphics[height=6.6cm]{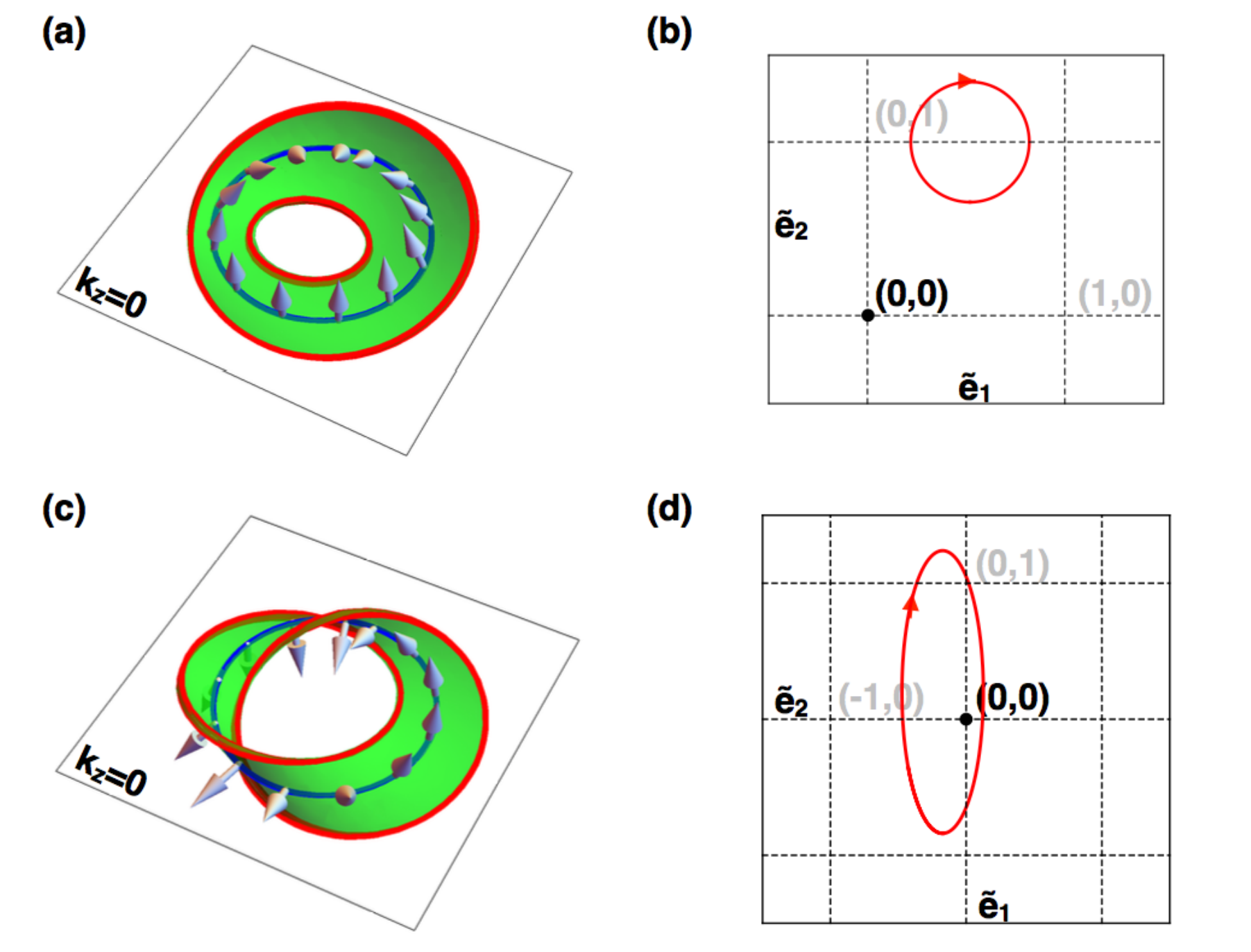}}
\caption{The winding number defined by $(\sqrt{3/4}g_z,2g_x)|_{{\bm k}_{NL}}$ along the blue nodal line. The left ones show the FSs (green), ELs (Red), nodal line (Blue) and normal vector of the FSs along the nodal line. The right ones show the corresponding winding numbers. The non-Hermitian perturbations with $\lambda=1$ in (a) and (b) is $g_x({\bm k})=0.2k_x+0.5$, $g_z({\bm k})=0.4k_y+0.3$ and $g_x({\bm k})=0.6k_x+0.1$, $g_z({\bm k})=0.4k_y-0.1$.
\label{F2}}
\end{figure}

{\em Winding number of Hopf-link EL semimetal---}The twisting property of the second equation in Eq. (\ref{n2}) discussed in the above section can be understood by a  winding number defined along the nodal line, as shown in Fig. \ref{F2}. We show that if the winding number is $\pm1$ (which is determined by the non-Hermitian perturbations), the Hopf-link ELs always emerge under sufficient weak perturbations ($\lambda\ll1$). On the other hand, if the winding number is $0$, the Hopf-link ELs never emerge.   
         As mentioned above, the observation of the winding number is the twisting structure of the FS as shown in Fig. \ref{F2} (a) and (c). This twisting property can be captured by the normal vectors of the FSs along the nodal line. To calculate the winding number, we first define the normal direction perpendicular to the FS, ${\bm d}({\bm k}_{nl})=2\lambda \nabla_{\bm k}{\bm h}({\bm k})\cdot {\bm g}({\bm k})|_{{\bm k}_{nl}}$, where we have used the fact ${\bm h}({\bm k}_{nl})=0$ and ${\bm k}_{nl}$ is the momentum on the nodal line. In order to simplify the discussion, we assume that the nodal line semiemtal has a mirror symmetry at the $k_z =0$ plan. The nodal line can be parameterized as $(k_{nl}^x(t),k_{nl}^y(t),0)$. The tangential direction of the nodal line is $(\partial_tk_{nl}^x(t),\partial_tk_{nl}^y(t),0)$. Using the coordinate system defined by ${\hat{\bm e}}_1\cdot (\hat{{\bm e}}_2\times\hat{\bm e}_3)=1$ with $\hat{\bm e}_1=N_1(\partial_tk_{nl}^x(t),\partial_tk_{nl}^y(t),0)$, $\hat{\bm e}_3=(0,0,1)$, where $N_1=1/\sqrt{(\partial_tk_{nl}^x(t))^2+(\partial_tk_{nl}^y(t))^2}$ is the normalization factor, we can obtain the $\hat{\bm e}_2$ vector, which is $\hat{\bm e}_2=N_1\{-\partial_tk_{nl}^y(t),\partial_tk_{nl}^x(t),0\}$. Projecting the vector ${\bm d}({\bm k}_{nl})$ onto $(\hat{\bm e}_2,\hat{\bm e}_3)$, we obtain, 
\begin{equation}
\tilde{{\bm d}}[{\bm k}_{nl}(t)]=(\tilde{d}_1[{\bm k}_{nl}(t)],\tilde{d}_2[{\bm k}_{nl}(t)]),
\end{equation}
where, 
\begin{equation}\begin{aligned}
&\tilde{d}_1[{\bm k}_{nl}(t)]=NN_1(-\partial_tk_{nl}^y(t)d_x[{\bm k}_{nl}(t)]+\partial_tk_{nl}^x(t)d_y[{\bm k}_{nl}(t)]),\\
&\tilde{d}_2[{\bm k}_{nl}(t)]=Nd_z[{\bm k}_{nl}(t)].
\end{aligned}\end{equation}
         Here $(d_x,d_y,d_z)$ are the three components of ${\bm d}[{\bm k}_{nl}(t)]$ and $N$ is the normalization factor of $\tilde{{\bm d}}[{\bm k}_{nl}(t)]$. Using the above two components of the $\tilde{{\bm d}}[{\bm k}_{nl}(t)]$ vector, the winding number can be defined as
\begin{equation}
\nu=\frac{1}{2\pi}\int_{-\pi}^{\pi}dt\varepsilon_{ij}\tilde{d}_i\partial_t\tilde{d}_j,
\end{equation}
where $\tilde{d}_{i}=\tilde{d}_1[{\bm k}_{nl}(t)],\tilde{d}_2[{\bm k}_{nl}(t)]$. 

In the example we discussed in the above section, $(k_{nl}^x(t),k_{nl}^y(t),0)=(\sqrt{3/4}\cos t,\sqrt{3/4}\sin t,0)$. We  obtain $\hat{\bm e}_1=(-\sin t,\cos t,0), \hat{\bm e}_2=(-\cos t,-\sin t,0)$. From ${\bm h}({\bm k})=(k_z,0,3/8-k^2/2)$ and ${\bm g}({\bm k})=(g_x,g_y,g_z)$, the vector of ${\bm d}({\bm k}_{nl})$ can be calculated as ${\bm d}({\bm k}_{nl})=2\lambda(-k_xg_z,-k_yg_z,g_x-k_zg_z)$. Notice $k_{nl}^z(t)=0$ and projecting onto the $(\hat{\bm e}_2,\hat{\bm e}_3)$ direction, we obtain 
\begin{equation}\begin{aligned}
\tilde{{\bm d}}[{\bm k}_{nl}(t)]&=2\lambda N(\sqrt{3/4}g_z[{\bm k}_{nl}(t)],g_x[{\bm k}_{nl}(t)]),
                     \label{S22}
\end{aligned}\end{equation}
         where $N$ is the normalized factor. The winding number is independent of the $g_y({\bm k})$ term due to the vanishing of $h_y({\bm k})$ in our model. The $g_y({\bm k})$ term only controls the width of the FS. According to Eq. (\ref{n2}),  the two ELs can not touch together at the nodal line if $|{\bm g}({\bm k}_{nl})|>0$. As a result, from Eq. (\ref{S22}), if the winding number is nonzero, $|{\bm g}({\bm k}_{nl})|$ is larger than zero so that  the two lined ELs can not touch together at the nodal line. Thus Hopf-link ELs always emerge with sufficient small value of $\lambda$ if the winding number is $\pm1$, and is totally independent of $g_y({\bm k})$ term in our model. Fig. \ref{F2}(b) and (d) show this winding number for the following two cases, $g_x({\bf k})=0.2k_x+0.5$, $g_z({\bf k})=0.4k_y+0.3$ and $g_x({\bf k})=0.6k_x+0.1$, $g_z({\bf k})=0.4k_y-0.1$. It is clear that when the winding number is nonzero, the FSs gain a twisting structure with Hopf-link ELs at the boundary.

{\em Discussion---}The bulk-boundary correspondence for the non-Hermitian systems has been widely discussed in the recent papers\cite{Alvarez,Wang2018a,Wang2018b,Kunst,Lee,Xiong,Nori}. As revealed by Ref\cite{Wang2018a,Wang2018b,Kunst}, the surface states of non-Hermitian systems may be different from  the one predicted by the Bloch Hamiltonian due to the non-Hermitian skin effect\cite{Wang2018a,Wang2018b,Kunst}. A complex-valued wavevector\cite{Wang2018a,Wang2018b} is introduced to obtain the correct bulk-boundary correspondence. The calculation of the surface states in our model will be discussed in a forthcoming paper. 


In summary, we have proposed a method to realize the Hopf-link EL semimetals from nodal line semimetals. The FSs of the Hopf-link EL semimetals have a twisting structure, whose boundaries are the Hopf-link. The linking number is constructed from a dual Hermitian Hamiltonian, whose nodal lines have the same topology with the ELs. We show that the Hopf-link phase is robust and can also be characterized by the winding number from the non-Hermitian perturbations along the nodal line.

{\em Note added}---During the preparation of this manuscript, we became aware of related work by Carlstrom, {\em et. al}\cite{Note added}.

{\em Acknowledgements}---We acknowledge helpful discussions with Dong. E. Liu and B. Andrei Bernevig. The work is supported by the Ministry of Science and Technology of China 973 program (No. 2015CB921300, No.~2017YFA0303100), National Science Foundation of China (Grant No. NSFC-1190020, 11534014, 11334012), and the Strategic Priority Research Program of CAS (Grant No.XDB07000000). QHW also acknowledges the supports by the NSFC funding No.11574134.

 \end{bibunit}
 \end{document}